\documentclass[journal,10pt]{IEEEtran}
\usepackage{amsmath,amsfonts,amssymb}
\usepackage{algorithm}
\usepackage{algpseudocode}
\usepackage{array}
\usepackage{textcomp}
\usepackage{stfloats}
\usepackage{url}
\usepackage{verbatim}
\usepackage{graphicx}
\usepackage{cite}
\usepackage{multirow}
 \usepackage{color}
\usepackage[table]{xcolor}
\usepackage{caption}
\usepackage{subcaption}
\usepackage{graphicx}

\usepackage{makecell}

\begin{document}
\title{Transceiver Design and Performance Analysis for LR-FHSS-based Direct-to-Satellite IoT}

\author{Sooyeob Jung, Seongah Jeong, Jinkyu Kang, Joon Gyu Ryu, and Joonhyuk Kang
\vspace{-25pt}
\thanks{This work was supported by Institute of Information \& communications Technology Planning \& Evaluation (IITP) grant funded by the Korea government (MSIT) (No.2020-0-00843, Development of low power satellite multiple access core technology based on LEO cubesat for global IoT service).}

\thanks{Sooyeob Jung is with KAIST and ETRI, Daejeon, South Korea (Email: jung2816@kaist.ac.kr), Seongah Jeong is with KNU, Daegu, South Korea (Email: seongah@knu.ac.kr), Jinkyu Kang is with Myongji Uni., Gyeonggi-do, South Korea (Email: jkkang@mju.ac.kr), Joon Gyu Ryu is with ETRI, Daejeon, South Korea (Email: jgryurt@etri.re.kr), and Joonhyuk Kang is with KAIST, Daejeon, South Korea (Email: jhkang@ee.kaist.ac.kr).}}

\markboth{IEEE Communications Letters}
{}

\maketitle

\begin{abstract}
This paper presents a novel transceiver design aimed at enabling Direct-to-Satellite Internet of Things (DtS-IoT) systems based on long range-frequency hopping spread spectrum (LR-FHSS). Our focus lies in developing an accurate transmission method through the analysis of the frame structure and key parameters outlined in Long Range Wide-Area Network (LoRaWAN) [\ref{Previous1}]. To address the Doppler effect in DtS-IoT networks and simultaneously receive numerous frequency hopping signals, a robust signal detector for the receiver is proposed. We verify the performance of the proposed LR-FHSS transceiver design through simulations conducted in a realistic satellite channel environment, assessing metrics such as miss detection probability and packet error probability.
\end{abstract}

\vspace{-0.2cm}

\IEEEpeerreviewmaketitle

\vspace{-0.3cm}
\section{Introduction}
Currently, there is ongoing progress in the development of low-power wide area network (LPWAN) technologies [\ref{LPWAN}] to facilitate satellite-based communication services. However, the existing LPWAN technologies utilizing low-Earth-orbit (LEO) satellites, such as Long Range (LoRa) [\ref{LoRa}] and Ingenu [\ref{Ingenu}], have demonstrated limitations in connecting remote areas. These technologies suffer from performance degradation caused by the Doppler effect and interference in LEO satellite channels [\ref{ETRIJ}], [\ref{IoT}]. Additionally, the LPWAN techniques with low network capacity are inadequate for satellite communications that require coverage over millions of square kilometers.

Recently, Semtech [\ref{Previous6}], a sponsor member of the LoRa Alliance [\ref{Previous1}], introduced a new uplink transmission technology for Direct-to-Satellite IoT (DtS-IoT) communications [\ref{DTS1}], [\ref{DTS2}]. This technology, known as long range-frequency hopping spread spectrum (LR-FHSS), employs a fast frequency hopping scheme to enhance network capacity and collision robustness in long-range and large-scale communication scenarios. Despite the limited availability of information on the LR-FHSS physical layer in the Long Range Wide-Area Network (LoRaWAN) specifications [\ref{Previous1}], a few research studies [\ref{Previous2}]-[\ref{Previous5}] have been conducted. Previous works mainly focused on throughput and outage probabilities. For instance, [\ref{Previous2}] compared the network capacity performance of LR-FHSS and LoRa under no fading channel conditions, while [\ref{Previous3}] mathematically investigated the success probability for the packet delivery under noise-free channel conditions, including path-loss and Rician fading in LR-FHSS systems. [\ref{Previous4}] derived a closed-form expression for outage probability under realistic channel conditions, considering path-loss and Nakagami-m fading. Furthermore, [\ref{Previous5}] explored a shadowed-Rician fading model, resembling the actual satellite channel, to analyze the outage probability of device-to-device-aided LR-FHSS schemes. The existing studies [\ref{Previous2}]-[\ref{Previous5}] demonstrate the suitability of LR-FHSS for DtS-IoT transmission, showcasing significant capacity improvements compared to LPWAN transmission schemes. However, none of these works have addressed the detailed physical-layer transmission and reception techniques specific to LR-FHSS, which remain an open challenge. Additionally, given that LR-FHSS transmission utilizes numerous frequency hopping blocks for interference management, accurate detection of the transmission-related header, such as the frequency hopping pattern for collision avoidance among multiple end devices (EDs), is crucial for the receiver. Robust signal detection becomes even more vital in the presence of the severe Doppler effect, a major performance degradation factor in LEO satellite communications.

In this paper, we develop a novel transceiver design for LR-FHSS-based DtS-IoT systems. In particular, the transmitter structure is designed based on the physical-layer specifications with the proposed time-on-air (TOA) calculation, which can be obtained by analyzing the frame structure and main parameters presented in LoRaWAN standard [\ref{Previous1}]. For the receiver, a signal detector is proposed to be capable of simultaneously receiving a huge number of frequency hopping signals, whose performances are verified to be compatible for the best sensitivity in Semtech’s application note [\ref{Previous6}] via simulations under the actual LEO satellite channel environment in terms of miss detection probability and packet error probability (PER).

\vspace{-0.2cm}
\section{Transceiver Design of LR-FHSS}\label{LR-FHSS}
\vspace{-0.1cm}
In this section, we briefly provide the main specifications of LR-FHSS focusing on the physical-layer features, and then detail the proposed transceiver design.

\vspace{-0.4cm}
\subsection{Overview of LR-FHSS}

\begin{table}[t]
    \caption{Main parameters of LR-FHSS.}\label{Table1}
    \centering
    \begin{tabular}{|>{\centering}m{1.75cm}|>{\centering}m{0.6cm}|>{\centering}m{0.6cm}|>{\centering}m{0.6cm}|>{\centering}m{0.6cm}|>{\centering}m{0.75cm}|>{\centering}m{0.75cm}|}
     \hline
     Region & \multicolumn{4}{c|}{EU863-870 MHz} & \multicolumn{2}{c|}{US902-928 MHz} \tabularnewline
     \hline
     LoRaWAN DR & DR8 & DR9 & DR10 & DR11 & DR5 & DR6 \tabularnewline
     \hline
     $N_{C}$ & 7 & 4 & 7 & 4 & 8 & 8 \tabularnewline
     \hline
     OCW [kHz] & \multicolumn{2}{c|}{137} & \multicolumn{2}{c|}{336} & \multicolumn{2}{c|}{1523} \tabularnewline
     \hline
     OBW [Hz]& \multicolumn{6}{c|}{488} \tabularnewline
     \hline
     Grid [kHz]& \multicolumn{4}{c|}{3.9} & \multicolumn{2}{c|}{25.4} \tabularnewline
     \hline
     $N_{CF}$ & \multicolumn{2}{c|}{280 (8x35)} & \multicolumn{2}{c|}{688 (8x86)} & \multicolumn{2}{c|}{3120 (52x60)} \tabularnewline
     \hline
     $N_{CF/ED}$ & \multicolumn{2}{c|}{35} & \multicolumn{2}{c|}{86} & \multicolumn{2}{c|}{60} \tabularnewline
     \hline
     $r$ & 1/3 & 2/3 & 1/3 & 2/3 & 1/3 & 2/3 \tabularnewline
     \hline
     Bit rate [bps] & 162 & 325 & 162 & 325 & 162 & 325 \tabularnewline
     \hline   
     ${N_H}$ & 3 & 2 & 3 & 2 & 3 & 2 \tabularnewline
     \Xhline{3\arrayrulewidth}
     \multicolumn{1}{!{\vrule width 1pt}c!{\vrule width 0.5pt}} {Max. $L$ [bytes]} & {58} & {123} & {58} & {123} & {58} & \multicolumn{1}{c!{\vrule width 1pt}}{133}\tabularnewline
     \hline   
      \multicolumn{1}{!{\vrule width 1pt}c!{\vrule width 0.5pt}} {Max. ${N_F}$} & {31} & {32} & {31} & {32} & {31} & \multicolumn{1}{c!{\vrule width 1pt}}{34} \tabularnewline
     \hline
      \multicolumn{1}{!{\vrule width 1pt}c!{\vrule width 0.5pt}} {${\rm{TO}}{{\rm{A}}_P}$ [ms]} & {3874.8} & {3743.7} & {3874.8} & {3743.7} & {3874.8} & \multicolumn{1}{c!{\vrule width 1pt}}{3948.5}
     \tabularnewline
     \Xhline{3\arrayrulewidth} 
\end{tabular}
\end{table}

The several features of LR-FHSS has been partially disclosed in LoRaWAN specification [\ref{Previous1}] published by LoRa Alliance and Semtech's application note [\ref{Previous6}], which are summarized in Table I. In [\ref{Previous1}], LoRaWAN data rate (DR) tables are sorted by region and frequency band, where LR-FHSS is especially defined in EU863-870 MHz and US902-928 MHz. In each region, the number of channels $N_C$, operating channel width (OCW) and grid are decided by considering the frequency plan with the fixed occupied bandwidth (OBW) of 488 Hz. Here, the grid is the minimum spacing between hopping channels. Using these parameters, the number of channels available for frequency hopping can be obtained as $N_{CF}={{{\rm{OCW}}} \mathord{\left/
 {\vphantom {{{\rm{OCW}}} {{\rm{OBW}}}}} \right.
 \kern-\nulldelimiterspace} {{\rm{OBW}}}}$, and the number of channels for frequency hopping per ED can be calculated as ${{{N_{CF/ED}} = {\rm{OCW}}} \mathord{\left/
 {\vphantom {{{N_{CF/ED}} = {\rm{OCW}}} {{\rm{grid}}}}} \right.
 \kern-\nulldelimiterspace} {{\rm{grid}}}}$. In addition, the bit rate, the number of header replicas ${N_H}$ and the maximum payload size $L$ are determined according to a coding rate $r$. The access to the channels is regulated by regional duty cycles, which constrain the TOA. Compared to the previous studies [\ref{Previous2}], [\ref{Previous3}] that provide the incorrect TOA of LR-FHSS packet resulting from the analytic errors in the construction of payload fragments, in this paper, we develop the correct TOA formula, and is given as
 \begin{align}
 {\rm{TO}}{{\rm{A}}_P} = {N_H}{T_H} + {N_F}{T_F}, \label{TOA}
 \end{align}
 where ${T_H}$, ${N_F}$ and ${T_F}$ are the header duration of 233.472 ms, the number of payload fragments and the payload fragment duration of 102.4 ms, respectively. Here, ${N_F}$ is calculated as
\begin{align}
{N_F} = \left\lceil {{N_{coded}}} \mathord{\left/
 {\vphantom {{{N_{coded}}} 48}} \right.
 \kern-\nulldelimiterspace} 48 \right\rceil  \simeq \left\lceil {{(L + 3)} \mathord{\left/
 {\vphantom {{(L + 3)} {N_{inf}}}} \right.
 \kern-\nulldelimiterspace} {N_{inf}}} \right\rceil,
\end{align}
where ${N_{coded}}$ is the number of coded payload bits, which is derived as ${N_{coded}} = {{\left( {8\left( {L + 2} \right) + 6} \right)} \mathord{\left/
 {\vphantom {{\left( {8\left( {L + 2} \right) + 6} \right)} r}} \right.
 \kern-\nulldelimiterspace} r}$ consisting of payload of $L$ bytes, CRC of 2 bytes and encoding tail bits of 6 bits, and ${N_{inf}}$ is the number of information bytes of payload fragment, which has the value of 2 or 4, when $r$ is 1/3 or 2/3, respectively. The proposed values of $N_F$ and ${\rm{TO}}{{\rm{A}}_P}$ are presented in Table I with ``bold box". The difference between TOAs in \eqref{TOA} and the previous works ranges up to 3306 ms in the case of DR5, by which the outage probability performance can be degraded by increasing the average number of interfering devices. For better understanding about the proposed TOA in \eqref{TOA}, we introduce the packet generation process.

\begin{figure}
\centering
\begin{subfigure}{0.35\textwidth}
    \includegraphics[width=\textwidth]{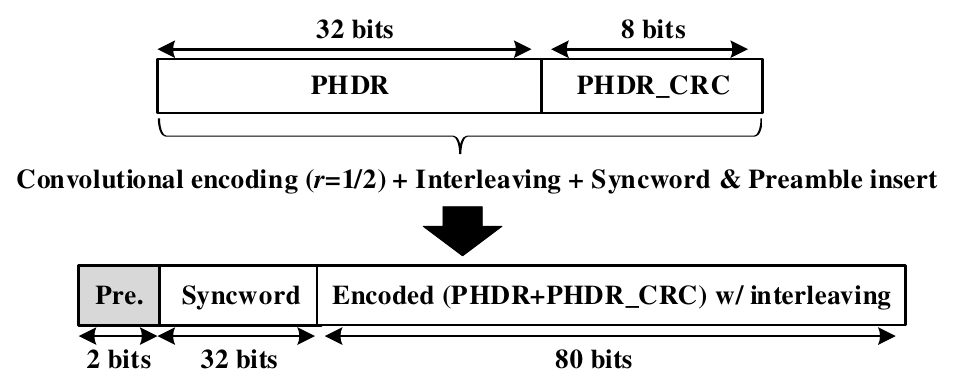}
    \caption{}
\end{subfigure}
\hfill
\begin{subfigure}{0.4\textwidth}
    \includegraphics[width=\textwidth]{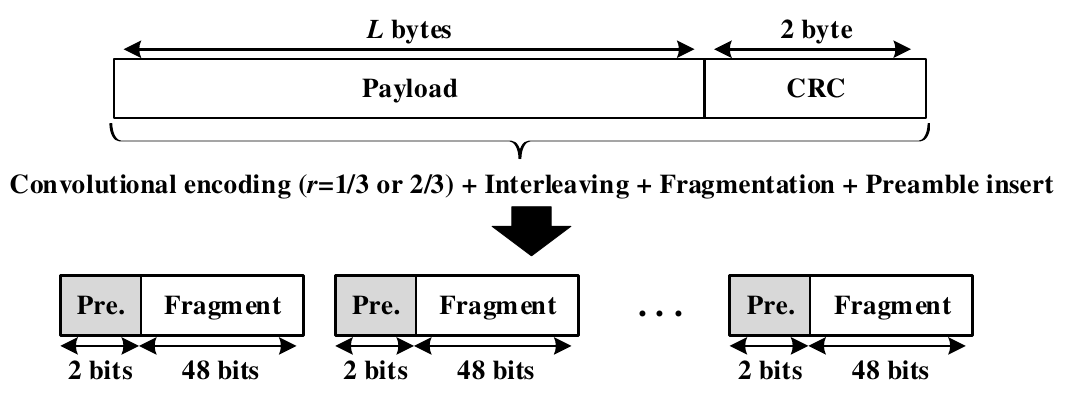}
    \caption{}
\end{subfigure}
\caption{Structure of (a) header and (b) payload fragments.}
\end{figure}

\begin{figure*}[t]\label{Tx}
\centering
    \includegraphics[width=13.5cm]{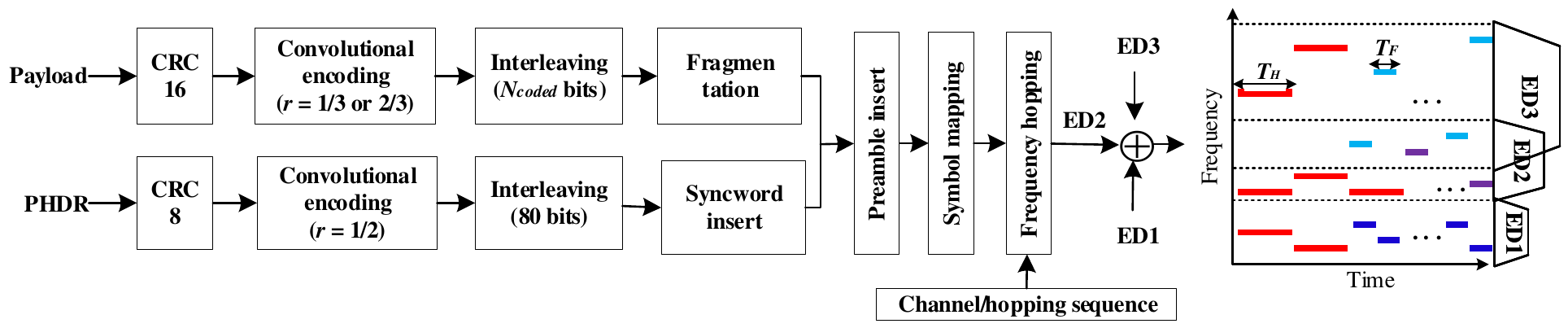}
    \caption{Transmitter structure of LR-FHSS with three EDs.}
\end{figure*}

\begin{figure*}[t]\label{Tx}
\centering
    \includegraphics[width=13.5cm]{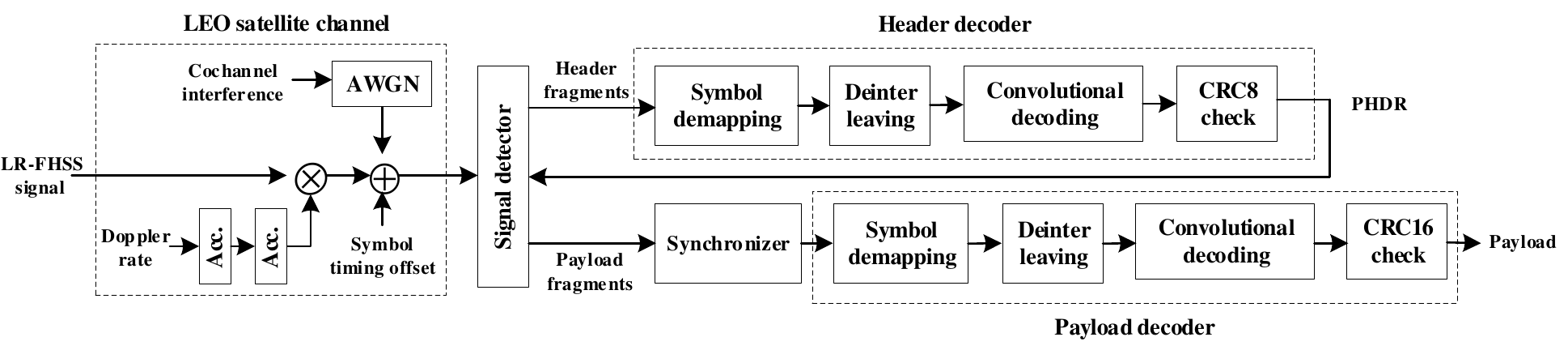}
    \caption{LEO satellite channel modeling and receiver structure of LR-FHSS.}
\end{figure*}

Fig. 1 shows the structure of the header and payload fragments of LR-FHSS. The PHY header (PHDR) consisting of the total of 32 bits contains the LR-FHSS transmission parameters [\ref{Previous2}]. After constructing the 32 bits PHDR and the following 8 bits $\rm{PHDR\_CRC}$ for CRC check, the convolutional encoding and interleaving are sequentially performed. Then, 2 bits for the preamble and 32 bits for the syncword are added to the front of the encoded (PHDR+$\rm{PHDR\_CRC}$) with interleaving as shown in Fig. 1(a). The payload of LR-FHSS is divided into the multiple fragments for frequency hopping by the payload fragmentation process as in Fig. 1(b). After performing the convolutional encoding with $r$ = 1/3 or 2/3 and interleaving on the payload and CRC sequentially, the resulting bits are fragmented into 48 bits each. By adding 2 bits of preamble to each fragment, the final payload fragment becomes 50 bits. Meanwhile, the existing works [\ref{Previous2}], [\ref{Previous3}] consider only one preamble to be added before the entire payload, which does not conform to the current LR-FHSS standard, and shortens the idle period for frequency hopping. This restricts the frequency offset tracking for each hopping block, making it difficult to implement a receiver in practice.

\vspace{-0.25cm}
\subsection{Transmitter Design of LR-FHSS}
Here, we explore the transmitter design of LR-FHSS with the frequency hopping. Fig. 2 shows the structure of LR-FHSS transmitter including the generation blocks of the header and payload fragments, by which the LR-FHSS packets from EDs are conveyed simultaneously in frequency-time domain. The PHDR data is generated via the several blocks consisting of CRC8 block, convolutional encoding block, interleaving block and syncword insertion block as shown in Fig. 1(a). As in Fig. 1(b), the payload fragments result from the combination of CRC16 block, convolutional encoding block, interleaving block and fragmentation block. The 112 bits of header and the 48 bits of payload fragments are finally composed of each block that performs frequency hopping transmission by inserting a 2 bits of preamble. The symbol mapping schemes for Gaussian minimum shift keying (GMSK) or quadrature phase shift keying (QPSK) can be applied to each hopping block, resulting in hopping blocks of 114 symbols for header and 50 symbols for payload data of GMSK or 57 symbols for header and 25 symbols for payload data of QPSK, respectively. In particular, GMSK is resistant to inter-channel interference due to its sharp spectral characteristics, and can enhance the signal detection performance due to the long symbol length. In the LR-FHSS signal output, the total of $N_H + N_F$ hopping blocks, considering the replicas of PHDR, are transmitted by switching frequency channels based on the hopping channel and hopping pattern, which are generated in the channel/hopping sequence block. In Fig. 2, the outputs of the LR-FHSS transmitters with the frequency hopping blocks are presented. In three EDs, two/three headers and several payload fragments are transmitted using frequency hopping, where the headers are placed at the start of the transmission with the longer duration in the clean frequency band. In the LR-FHSS networks, the collision rate between the multiple EDs can be reduced by managing the spectrum used by each ED. Specifically, the LR-FHSS networks need to allocate the clean spectrum for EDs with the high quality of service (QoS) or with low transmit power.

\vspace{-0.35cm}
\subsection{LEO Satellite Channel Modeling}
The LR-FHSS signal conveyed from the transmitter undergoes a LEO satellite channel model, such as the Doppler effect, symbol timing offset and noise. As shown in Fig. 3, we consider the following channel models for the transceiver implementation. In the LEO satellite communication systems, a large and time-variant frequency shift can be observed due to the Doppler effect depending on the carrier frequency and the characteristics of the assigned LEO satellite such as altitude, orbit and coverage [\ref{ETRIJ}]. For realizing the LEO satellite channel environments, the Doppler effect is generated and applied as Doppler rate passes through two accumulators. In the proposed architecture, the symbol timing offset that may occur during the sampling process of the receiver is considered. Moreover, in the DtS-IoT systems, the unslotted Aloha-based multiple access is preferred in the way of increasing the network capacity, while it may induce the cochannel interference among the multiple EDs. The cochannel interference can be realized with noise by randomly selecting the overlapping ratio of hopping blocks of different EDs in the total frame length. However, in this paper, we assume the perfect synchronization, for which the synchronizer design of transceiver to cope with the cochannel interference remains as our future work.

\vspace{-0.35cm}
\subsection{Receiver Design of LR-FHSS}
In this section, we develop the receiver design of LR-FHSS, which enables to simultaneously receive the LR-FHSS packets. The receiver is composed of the signal detector, synchronizer, header decoder and payload decoder illustrated in Fig. 3. In particular, the LR-FHSS signal detector needs to be designed for simultaneous reception of the LR-FHSS packets from multiple EDs, which is described in detail in the following section. From the header position information obtained in the signal detector, the header fragments can be transferred from the external memory to header decoder. The header decoder involves with the symbol demapping block, the deinterleaving block, the convolutional decoding block and the CRC8 check block, and finally outputs the PHDR. Using the PHDR, the payload fragments can be transferred from the external memory to synchronizer, and then the LR-FHSS signal subjected to satellite channel effects can be compensated in the synchronizer. From the synchronized payload fragments, we can extract the payload data through the payload decoder.

\vspace{-0.12cm}
\section{Signal Detector Development for LR-FHSS}\label{sec:detect}
Here, we propose the detailed structure of LR-FHSS signal detector in Fig. 4, which is composed of channelizer, external memory and header detector.


\begin{figure}[t]\label{SD}
    \includegraphics[width=\columnwidth]{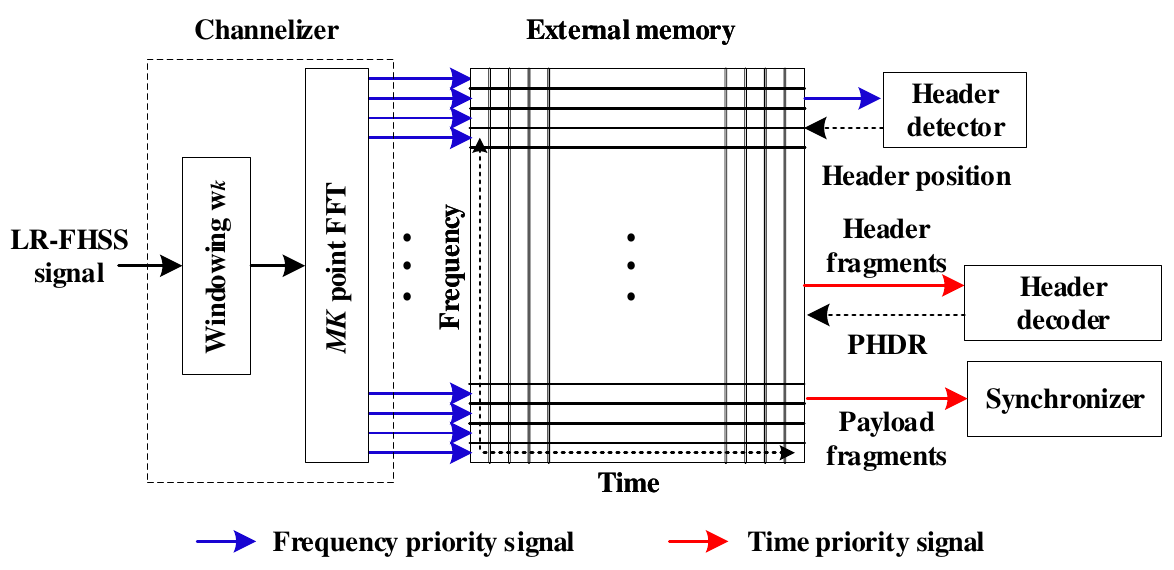}
    \caption{Signal detector structure of LR-FHSS.}
\end{figure}

\vspace{-0.3cm}
\subsection{Channelizer}
At the receiver, the LR-FHSS signal firstly experiences the channelizer in order to classify the frequency hopping channels by the FFT operation with windowing, the channelizer has the structure that can simultaneously receive up to $N_{CF} = 3120$ signals available in $\rm{OCW} = 1523 \rm{kHz}$. The sampled signal of LR-FHSS, which is oversampled by a factor of 8, is used to detect the frequency hopping signals. For example, let us denote the LR-FHSS signal as $s\left( t \right) = A\left( t \right){e^{j\left( {2\pi {f_c}\left( t \right)} \right)}},$ where $A\left( t \right)$ denotes the signal in baseband, and ${{f_c}\left( t \right)}$ denotes the time-varying carrier frequency. The frequency hopping of the LR-FHSS signal is performed by changing the carrier frequency ${{f_c}\left( t \right)}$ according to the channel/hopping sequence at certain time intervals, two types of which are defined as $T_H$ and $T_F$ interval for header and payload fragments, respectively. The sampled signal at the receiver can be then modeled as ${{\bf{r}}_k} = {{\bf{s}}_k} + {{\bf{u}}_k},$ where ${{\bf{u}}_k}$ is the additive white Gaussian noise (AWGN). By considering the $M$ channels distributed at grid intervals within OCW with the $K$ FFT bins assigned to each channel, the FFT operation of the length $MK$ is used to compute the power spectral estimates for the $M$ channels. After the windowing the vectors ${{\bf{r}}_k}$ with the symmetric window ${{\bf{w}}_k} = \left[ {{w_0},{w_1}, \cdots ,{w_{N - 1}}} \right]$ of length $N$ for low pass filtering is performed, the result of FFT operation can be given as ${{{X}}_m} = \sum\nolimits_{k = 0}^{N - 1} {{w_k}{r_k}{e^{\frac{{ - j2\pi km}}{MK}}}}$. The input data of 8 oversamples is downsampled to two oversamples in windowing. Finally, the FFT output of the channelizer is transferred to the external memory per frequency bin.

\vspace{-0.3cm}
\subsection{External Memory}
The external memory followed by the channelizer is divided into the memory regions, in which the data is stored based on the frequency axis and the time axis. The FFT output corresponds to all $MK$ frequency signals at one sampling point, which is referred to as frequency priority signal. In the time domain, two oversampled signals for each frequency are stored in the time order, which is called a time priority signal. First, the frequency priority signal and the time priority signal are sequentially stored in the corresponding memory region, respectively. To detect the header position, the frequency priority signal for each frequency is transmitted to the header detector, which makes a detection decision by using the cross-correlation. The detection frequency and time determined by the header detector become the header hopping channel information and the header reception time information, respectively. The header position information obtained from the header detector is transmitted to the external memory, while the time priority signal of the header fragments located at a sampling point of the corresponding frequency channel is transferred to the header decoder. From the PHDR by header decoding, we can find out the 9 bits channel/hopping sequence, which contains the hopping channel information of the payload fragments, that is, the entire payload fragments can be collected and then passed to synchronization block.

\vspace{-0.2cm}
\subsection{Header Detector}
The header detector based on FFT and windowing can obtain the header position information, which consists of a frequency hopping channel and a sampling point. The frequency priority signal from the external memory is windowed with the size $N$ for cross-correlation. The cross-correlation is performed with the filter coefficient ${{\bf{c}}_k} = \left[ {{c_0},{c_1}, \cdots ,{c_{N - 1}}} \right]$ based on the synchronous word of hexadecimal 0x2C0F7995. The result of $M$ point FFT operation with windowing of length $N$ can be written as ${{{X}}_m} = \sum\nolimits_{k = 0}^{N - 1} {{c_k}{r_k}{e^{\frac{{ - j2\pi km}}{M}}}}$, from which we can derive the power spectral estimate ${{{P}}_m} = {\left| {{{{X}}_m}} \right|^2}$. By finding the peak value of ${{{P}}_m}$ in time, we can estimate the header position corresponding to the sampling point and the coarse carrier frequency offset (CFO) corresponding to the frequency hopping channel. In addition, the fine CFO can be estimated with the resolution of ${{976} \mathord{\left/
 {\vphantom {{976} M}} \right.
 \kern-\nulldelimiterspace} M}$ Hz by finding the peak value of ${{{P}}_m}$ in frequency. In the proposed signal detector, the peak value found in the power spectral estimates can be defined as the header position if no larger peak occurs during a certain sample period. If the peak search interval is short, the signal detection accuracy increases with the high resolution. However, a large number of signal detection results significantly increases the receiver complexity. Finally, the header position information is transmitted to the external memory for header decoding and synchronization.

\vspace{-0.25cm}
\section{Simulation Results}\label{sec:simul}
In this section, we evaluate the performance of the proposed transceiver design of LR-FHSS in terms of miss detection probability and PER. The miss detection probability is measured through the estimated header position, and the header PER and payload PER is measured using the results of header decoding block and payload decoding block, respectively. In the header decoding block, we apply the joint carrier sensing and header reception method that is well-known to be robust against Doppler rate, and minimizes the miss detection by allowing false to be filtered out using the CRC of the header. By referring [\ref{Previous5}], the parameters for simulations are summarized as follows: $f_c = 900 \rm{MHz}$, $\rm{OCW} = 39.06 \rm{kHz}$, $\rm{Grid} = 3.9 \rm{kHz}$, $N_{CF} = 80$, $N_{CF/ED} = 10$, $r = 1/3$, $L = 32 \rm{bytes}$, $N_H = 1$ and $N_F = 18$. The miss detection probability is measured by Monte Carlo simulation with iteration of 1000, and the PER performance is compared with the reference method in [\ref{Previous6}]. The proposed signal detector is developed with the following parameters: the windowing length $N = 16$, FFT length $M = 4096$, FFT bins $K=2$ for channelizer and the windowing length $N=32$ (QPSK) or $64$ (GMSK), FFT length $M=128$ for header detector. In the header detector, the use of GMSK can have excellent performance because it can use twice the symbol length per hopping block of QPSK. Also, the GMSK with a $BT$ factor equal to 0.3 is considered.

Fig. 5 shows the miss detection probability of signal detector with GMSK and QPSK according to the different symbol timing offset, peak search period and Doppler rate. In Fig. 5(a), the miss detection probability increases as the peak search interval increases to [12 24 48] bits in channel environment without Doppler rate. The long peak search interval can result in many peak values, which prevents the accurate signal detection. On the other hand, using the long peak search interval has the advantage of reducing implementation complexity by reducing the number of peak value searches. In the 48 bits peak search interval, as the Doppler rate increases to [0 200 400] Hz, the signal-to-noise ratio (SNR) deteriorates by about 1.5 dB. However, even in the environment with Doppler effect, the overall miss detection probability is low compared to the case of QPSK in Fig. 5(b). In all cases involving the peak search interval and the Doppler rate, symbol timing offset has no effect on the signal detection performance. This is because GMSK can tolerate the inter-symbol interference influence at the cost of spectral properties. Compared to the GMSK case, the performance degradation of QPSK-based LR-FHSS according to the peak search interval is the same. On the other hand, there is no change in the miss detection performance according to the Doppler rate in the 48 bits peak search interval. Also, when the symbol timing offset increases, a performance gap occurs in high SNR. Here, a symbol timing offset of 0 is considered equal to 4/8 since the receiver uses the two oversampled data. These results show that QPSK is more vulnerable to symbol timing offset than GMSK, but stronger to the Doppler effect. Overall, the miss detection probability of QPSK is 3 dB lower than that of GMSK due to the short symbol length. Accordingly, GMSK can be suitable for LR-FHSS transmission with a limited symbol length.

\begin{figure}[t]
     \centering
     \begin{subfigure}[b]{0.48\columnwidth}
         \centering
         \includegraphics[width=\columnwidth, height=\columnwidth]{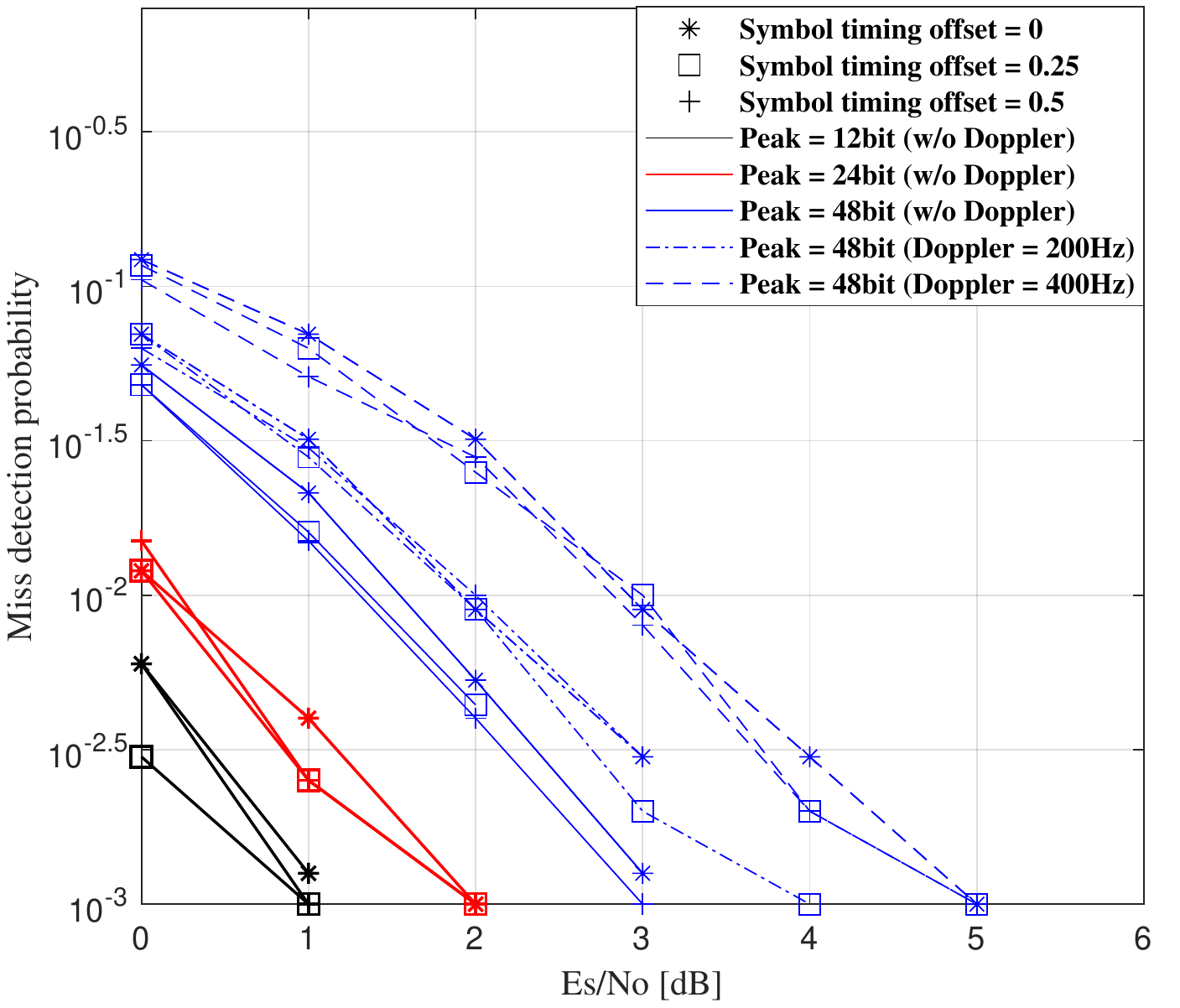}
         \caption{}
         \label{aa}
     \end{subfigure}
     \hfill
     \begin{subfigure}[b]{0.48\columnwidth}
         \centering
         \includegraphics[width=\columnwidth, height=\columnwidth]{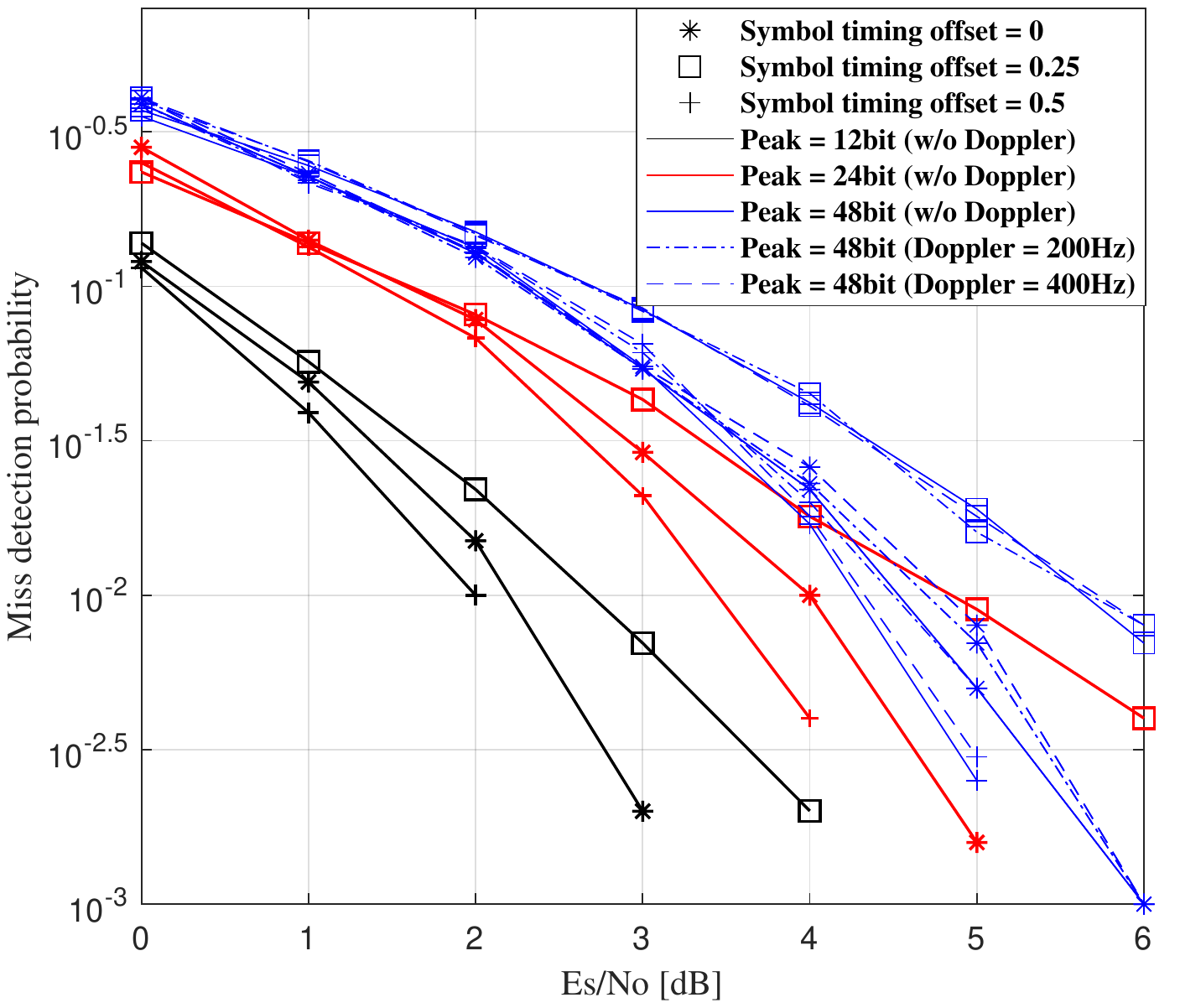}
         \caption{}
         \label{bb}
     \end{subfigure}
        \caption{Miss detection probability of (a) GMSK and (b) QPSK-based LR-FHSS signal detection under satellite environment.}
        \label{cc}        
\end{figure}

\begin{figure}[t]\label{PER}
\centering
    \includegraphics[width=0.29\textwidth]{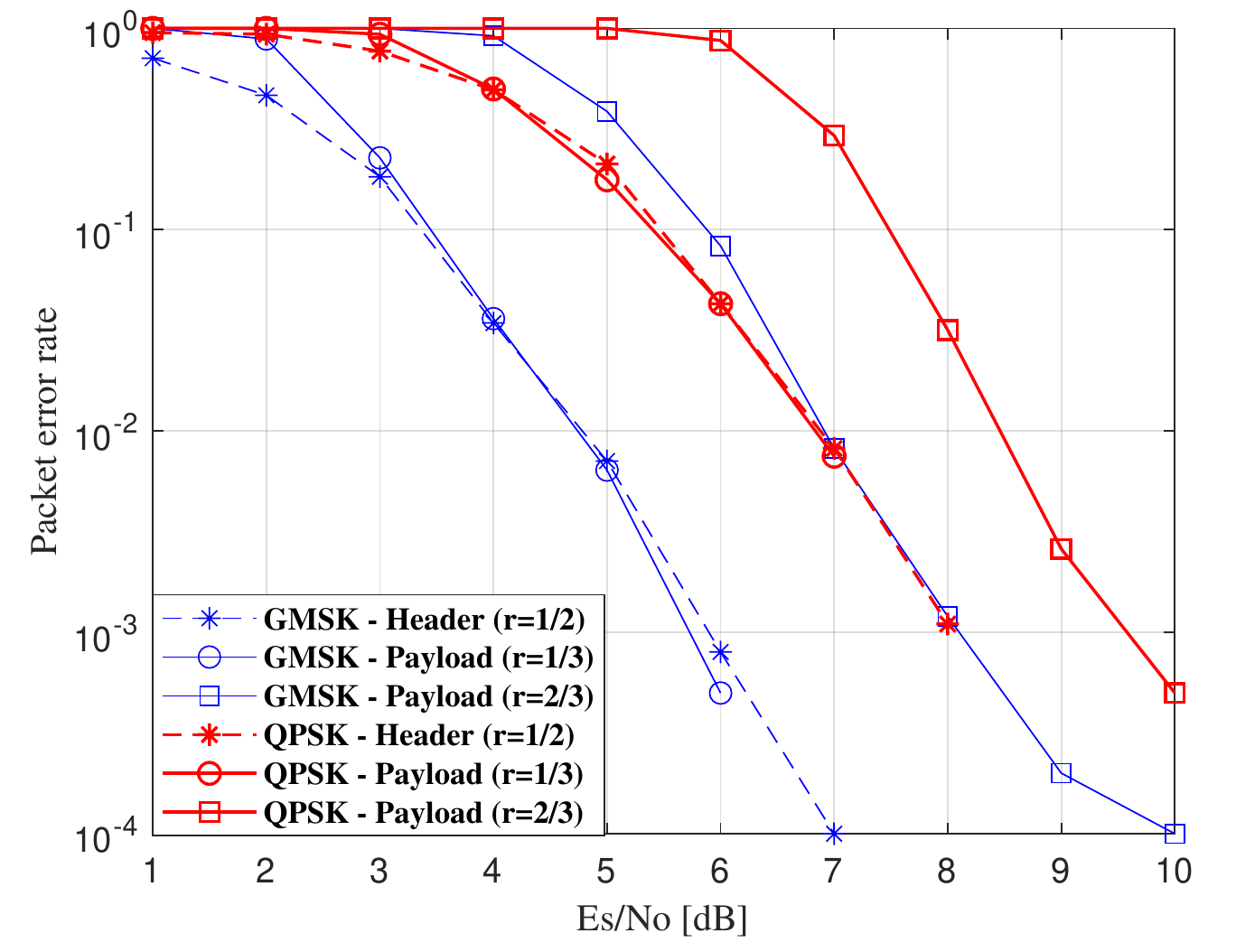}
    \caption{PER of LR-FHSS header and payload under AWGN.}
\end{figure}

Fig. 6 illustrates the header and payload PERs in GMSK and QPSK-based LR-FHSS transmission under AWGN with the perfect synchronization. In the proposed design, the signal detector can be verified by measuring the header PER instead of the miss detection probability because the header decoder allows the false detection in the decoding process. In the header PER of GMSK, the SNR of 6 dB is required to satisfy the PER of $10^{-3}$, which is equivalent to the required SNR obtained from the best sensitivity of -137 dBm in Semtech's application note [\ref{Previous6}]. For the required SNR, a noise figure of 6 dB, a bandwidth of 488 Hz and a noise power of 174 dBm are considered. It is verified in Fig. 6 that the proposed signal detector can achieve the reference performance provided by [\ref{Previous6}]. In addition, the PER of GMSK shows the SNR gain of about 2 dB over those of QPSK. Because of the header with the fixed code rate of 1/2, there is the limit in that the PER performance of a payload with the stronger code rate of 1/3 can not be lowered than the header PER. This limitation shows the room for improvement of the LR-FHSS performance, which remains as the future work.

\vspace{-0.25cm}
\section{Conclusions}\label{sec:con}
In this paper, we have proposed the LR-FHSS transceiver structure for DtS-IoT communication by analyzing the physical layer features based on the LoRaWAN specification [\ref{Previous1}]. In particular, we provide the detailed transmitter design with the proposed TOA calculation, while, for the receiver, the signal detector composed of a channelizer, an external memory and a header detector is developed to be robust against the Doppler effect as well as be capable of receiving a large number of frequency hopping signals simultaneously. Through simulations, it is verified that the LR-FHSS system based on GMSK modulation exhibits better performance in terms of miss detection probability and PER compared to the LR-FHSS system based on QPSK modulation. Additionally, this paper demonstrates that the header PER performance meets the standard requirements. As the future works, we suggest implementing an FPGA-based LR-FHSS transceiver that includes the synchronization block to compensate for the effects of LEO satellite channel.

\vspace{-0.25cm}
\bibliographystyle{IEEEtran}

\end{document}